\documentstyle[preprint,prl,aps]{revtex}
\begin{document}
\draft
\preprint{\vbox{
\hbox{IFP-465-UNC}
\hbox{hep-ph/9308275}
\hbox{August 1993}
}}

\title{
Cubic Mass Relation in
$\rm SU(3)\times U(1)$ Electroweak Theory
}

\author{Paul H. Frampton,
Plamen I. Krastev\footnote{Permanent address: Institute of Nuclear Research
and Nuclear Energy, Bulgarian Academy of Sciences, BG--1784 Sofia,
Bulgaria.}
and James T. Liu}
\address{
Institute of Field Physics, Department of Physics and Astronomy,\\
University of North Carolina, Chapel Hill, NC 27599--3255, USA
}
\maketitle
\begin{abstract}
{

In the 331 model, lepton number may be explicitly broken by trilinear scalar
self couplings.  This leads to neutrino masses proportional to the cube of
the corresponding charged lepton mass, with consequences for solar neutrinos
and for hot dark matter.
}
\end{abstract}
\newpage

\narrowtext

Whether neutrinos have non-zero rest mass is still an unresolved question.
It is possibly the question whose answer will provide the first evidence of
physics beyond the minimal standard model of particle interactions.
Theoretically, there is no known symmetry reason for vanishing neutrino
masses; experimentally, there are indications of non-vanishing neutrino
masses which await confirmation.  In the present letter we anticipate the
emergence of non-zero neutrino masses and derive a cubic relation
between neutrino and charged lepton masses.

Concerning the experimental situation, solar neutrino
experiments \cite{solar} suggest a
deficit of solar neutrinos compared to the predictions of the standard solar
model; if this is interpreted as due to the Mikheyev--Smirnov--Wolfenstein
(MSW)
mechanism \cite{msw} one is led to expect $m(\nu_\mu)\sim 10^{-3}$eV while
$m(\nu_e)$ is much smaller \cite{mswstat}.  Also, if $\nu_\tau$ is a
significant component
of hot dark matter in the universe one expects \cite{harari} $m(\nu_\tau)\simeq
30$eV.  Thirdly, the measurements of atmospheric $\nu_\mu$ and $\nu_e$
neutrinos suggest an anomalously small muon neutrino flux which may be
interpreted as due to non--zero neutrino masses of either $\nu_\mu$ or
$\nu_\tau$ of the order of $10^{-1}$eV \cite{atmo}.

These provocative hints from experiment lead inevitably to the question
of how neutrino masses are generated in the theory.
A popular conjecture is the see-saw mechanism \cite{seesaw}
which naturally accounts for why neutrino masses are small.  In
a {\it quadratic} see-saw, $m(\nu_i)$ is proportional to $m(f)^2/M$ where $f$
is some charged fermion in the same family and $M$ is a larger mass scale;
this is based on a grand
unification scenario with a desert hypothesis and on the assumption that
right-handed neutrino fields exist, one for each known fermion family.
It is usually assumed that $f$ is the up-quark member of the same family,
based on $SO(10)$ arguments (see {\it e.g.}~\cite{sogut} and references
therein).

Here we examine the question of neutrino masses in the 331 model
\cite{frampton,pleitez}.  In the minimal version of the 331
model there
are no right-handed neutrinos.  Furthermore, while lepton family number
is not conserved, total lepton number $L=L_e+L_\mu+L_\tau$ is conserved
in the minimal 331 model.  As a result, neutrinos remain massless just
as in the minimal standard electroweak model.

It is straightforward to give neutrinos a Dirac mass in the 331 model
by the introduction of right-handed neutrinos.  However this approach
has the same drawback as in the standard model, namely the Yukawa
couplings have to be unnaturally small in order to account for the
small neutrino masses.  We are thus led to consider the case where
neutrinos pick up a Majorana mass, which necessarily involves violating
lepton number.  Spontaneous breaking of $L$ will lead to a massless
triplet Majoron \cite{gelmini} which has been ruled out experimentally
by $Z$ lineshape measurements as mentioned briefly in \cite{frampton}.
Thus if $L$ is broken it must be done explicitly.

We first review the relevant features of the 331 model.  The electroweak
gauge group is ${\rm SU}(3)_L\times {\rm U}(1)_X$ and the leptons are
in antitriplets with $X=0$, $L_{i\alpha}=(\ell_i^-,\nu_i,\ell_i^+)$ for
$i=e,\mu,\tau$.  The quark sector is unimportant for the following
discussion, but, via anomaly cancellation, gives rise to the three
family requirement.  Symmetry breaking and fermion masses arise from
four Higgs multiplets, of which only two couple to the leptons.  These
are a sextet Higgs $H^{\alpha\beta}$ and a triplet Higgs $\phi^\alpha$
both with $X=0$.  When ${\rm SU}(3)_L\times {\rm U}(1)_X$ is broken
down to ${\rm SU}(2)_L\times {\rm U}(1)_Y$, the sextet decomposes as
${\bf 6}_0\to {\bf3}_2+{\bf2}_{-1}+{\bf1}_{-4}$ into SU(2) triplet,
doublet and singlet Higgs, and the triplet decomposes as ${\bf 3}_0\to
{\bf2}_{1} +{\bf1}_{-2}$ into a doublet and singlet.

The lepton Yukawa couplings may be written
\begin{equation}
-{\cal L} = h_1^{ij} L_\alpha'^i L_\beta'^j H^{\alpha\beta}
	+ h_2^{ij} L_\alpha'^i L_\beta'^j \overline{\phi}_\gamma
		\epsilon^{\alpha\beta\gamma} + \rm h.c.\ ,
\label{yukawa}
\end{equation}
where the primes denote weak eigenstates.  Here $h_1^{ij}$ is symmetric
in family space while $h_2^{ij}$ is antisymmetric.  In order for the
Yukawas to conserve $L$, the SU(2) triplet and singlet Higgs must carry
lepton numbers $-2$ and 2 respectively while the doublets have $L=0$.

In the minimal 331 model, charged lepton masses arise from the scalar VEVs
$\langle\phi^2\rangle=v/\sqrt{2}$ and $\langle H^{13}\rangle \equiv\langle
H^{31}\rangle=y/2$.  The resulting mass matrix is given in family space
by
\begin{equation}
m(\ell)^{ij}=h_1^{(ij)}y+\sqrt{2}h_2^{[ij]}v\ .
\end{equation}
Assuming $h_1^{ij}$ and $h_2^{ij}$ are real so there is no CP violation
in the leptonic sector, we may pick a basis where
$h_1^{ij}=h_1^i\delta^{ij}$ is diagonal and $h_2$ remains antisymmetric.
There are then a total of six Yukawa parameters: three charged lepton
masses and three mixing angles.  To avoid unsuppressed lepton flavor
violating processes \cite{glashow}, we demand $h_2v\ll h_1y$.  In this
case, we find charged lepton masses $m(\ell_i) \approx h_1^iy$ and
mixing angles $\alpha^{ij}\approx \sqrt{2}h_2^{ij}v/(h_1^i+h_1^j)y$.
In the minimal case where neutrinos remain massless, they can be
rotated so the $W^\pm$ charged current is lepton family diagonal.
However, the $\alpha^{ij}$ angles show up in the couplings to both
singly and doubly charged dileptons \cite{jtliu}.

We now give neutrinos masses by breaking $L$ explicitly \cite{pleitez2}.
This is done
in the Higgs sector by adding the two soft (dimension three) cubic self
couplings
\begin{equation}
-{\cal L} = m_1 H^{\alpha_1\beta_1}H^{\alpha_2\beta_2}H^{\alpha_3\beta_3}
	\epsilon_{\alpha_1\alpha_2\alpha_3}\epsilon_{\beta_1\beta_2\beta_3}
	+ m_2 H^{\alpha\beta}\overline{\phi}_\alpha\overline{\phi}_\beta
	+ \rm h.c.\ .
\label{cubic}
\end{equation}
Both terms violate $L$ by two units.  The cubic couplings in general
induce a linear term in $H^{22}$ and shifts the vacuum value to
$\langle H^{22}\rangle = (3m_1y^2-m_2v^2)/2M(H^{22})^2$.  This triplet
Higgs VEV is constrained by the $\rho$--parameter, but, if present,
gives rise to tree-level lepton number violation and Majorana neutrino
masses.  Since lepton number has been broken explicitly, there would be
no massless Majoron in this case.

An interesting case occurs when $3m_1y^2=m_2v^2$, which implies that at
tree level the vacuum value of $H^{22}$ vanishes.  Imposing this
requirement that $\langle H^{22}\rangle=0$ prevents large tree-level
lepton number violating processes and furthermore gives agreement with
the experimentally observed limits on the $\rho$--parameter.  Here we
shall further assume that $\langle H^{22}\rangle$ remains zero even
with radiative corrections.  This necessarily involves fine tuning order
by order in perturbation theory.  This assumption is made to obtain an
interesting new pattern of the neutrino masses which might be derived
more naturally in some future different model.

In the minimal 331 model, with $m_1=m_2=0$ in Eq.~(\ref{cubic}), lepton
number remains an exact symmetry of the theory.  In this case, with the
assumptions we have made (especially $\langle H^{22}\rangle=0$) the
neutrinos cannot acquire mass.  If we further set the antisymmetric
Yukawa coupling $h_2^{ij}$ to zero, then the Yukawa interactions can be
taken to be family diagonal.  As a result, only dileptons can violate
lepton family number, and hence there is the selection rule
$\Delta L_i=0$~mod~2.

Next consider turning on the soft lepton number violating terms,
$3m_1y^2=m_2v^2\ne0$ in Eq.~(\ref{cubic}), and keeping
$h_2^{ij}=0$ in Eq.~(\ref{yukawa}).  In this case, the lepton Yukawas remain
family diagonal, and we may work in the basis for the
$L_\alpha^i$ fields in which $h_1^{ij}$ is diagonal,
$h_1^{ij}=h_1^i\delta^{ij}$.  Since we have imposed $\langle
H^{22}\rangle=0$, the neutrinos are massless at tree level.
At one-loop level, however, the diagram of Fig.~\ref{fig1a}($a$)
gives a finite contribution to the neutrino masses.

Treating the lepton number violating mass insertion on the Higgs line,
$\delta m^2=6m_1y$, as a perturbation, the one-loop diagram is
evaluated as
\begin{eqnarray}
i\Sigma^i(p)&=&2\int\!{d^4k\over(2\pi)^4}\,(i2h_1^i)\gamma_L
{i\over \not\! k{}+\!\not\! p-m_i}(i2h_1^i)\gamma_L
{i\over k^2-M(H^{23})^2}(i\delta m^2) {i\over k^2-M(H^{12})^2}\nonumber\\
&=&-8(h_1^i)^2m_i\delta m^2\gamma_L\int\!{d^4k\over(2\pi)^4}\,
{1\over (k+p)^2-m_i^2} {1\over k^2-M(H^{23})^2}{1\over k^2-M(H^{12})^2}\ ,
\end{eqnarray}
where $i$ labels the lepton family and $\gamma_L={1\over2}(1-\gamma^5)$
projects onto left-handed chiralities.  The factor of 2 takes into account
both the diagram and the conjugate one.
It is straightforward to perform the loop integration in the limit $p\ll
m_i\ll M(H)$.  Using $m(\ell_i)=h_1^iy$, we arrive at the result
\begin{equation}
m(\nu_i)=C m(\ell_i)^3/M_W^2\ ,
\label{cubicmass}
\end{equation}
where the dimensionless constant $C$ is given by
\begin{equation}
C={\alpha_2\over2\pi\sin^2\beta} {\delta m^2\over M(H^{23})^2-M(H^{12})^2}
\ln\left({M(H^{23})^2\over M(H^{12})^2}\right) \ .
\label{constant}
\end{equation}
We have defined $\tan\beta=y/\sqrt{v^2+v'^2}$ where $v$, $v'$ and $y$ are
the VEVs of the three Higgs doublets giving mass to the $W^\pm$,
$M_W={g\over2}\sqrt{v^2+v'^2+y^2}$.

Eq.~(\ref{cubicmass}) is the cubic relationship between neutrino mass
and charged lepton mass.  Unlike the model of Ref.~\cite{zee} where the
various Yukawa couplings are unrelated, in the 331 model, the {\it
same} Yukawa couplings, Eq.~(\ref{yukawa}), are responsible for both
the charged lepton mass and, through the loop diagram, the neutrino
mass.  This is the origin of the above cubic mass relationship.  More
generally, this feature of the 331 model leads to definite structures of
both neutrino masses and mixing.

Family mixing in the 331 model is complicated both by the fact that there
are additional couplings to dileptons and by the large number of scalars in
the model.  In order to simplify the discussion, we work with small mixing
in the leptonic sector.  In this case, we can parametrize the charged lepton
mixing by the three angles $\alpha^{ij}$ as before.  Diagonalizing the
neutrino mass matrix gives rise to neutrino mixing angles $\theta^{ij}$.

While neutral currents are always family diagonal in the leptonic
sector, mixing angles in the charged currents are given
approximately by
$\alpha^{ij}-\theta^{ij}$ for the coupling to $W^-$,
$\alpha^{ij}+\theta^{ij}$ for $Y^-$ and $2\alpha^{ij}$ for
$Y^{--}$ (all in the limit of small mixing) \cite{jtliu}.  The
strongest constraints on $\alpha^{ij}$ come from the lack of
observation of lepton family violating decays such as $\mu\to3e$ whose
branching ratio is below $10^{-12}$ \cite{pdb}.  This process may occur via
tree-level $Y^{--}$ exchange in the 331 model.  Since the dilepton mass is
bounded to be under a TeV \cite{frampton,daniel}, this implies
$\alpha^{12}<10^{-5}$, which means that, in Eq.~(\ref{yukawa}),
$h_2^{12}v<10^{-5}h_1^{2}y$.  Somewhat weaker limits can also be placed
on first--third and second--third family mixing.

Since we have assumed $h_2^{ij}=0$ in deriving the cubic relation
above, only the first term of Eq.~(\ref{yukawa}) is present and the
Yukawa couplings can be made family diagonal.  As a result,
$\alpha^{ij}=\theta^{ij}=0$ and there is {\it no mixing} of neutrino or
charged lepton flavors.  In order to obtain mixing and $\Delta
L_i=\pm1$ lepton family violation we must make $h_2^{ij}\ne 0$ which
gives rise to non-diagonal mass matrices.

In this more general case, the neutrino mixing angles $\theta^{ij}$ are
not independent, but are related to the $\alpha^{ij}$ because of the
common Yukawa couplings.  When $h_2^{ij}\ne0$ the remaining graphs
($b$) through ($d$) depicted in Fig.~\ref{fig1a} also contribute to the
neutrino mass matrix.  In general, the relation between the mixing
angles depends on the ratio of the soft breaking masses $m_1$ and $m_2$
in Eq.~(\ref{cubic}) as well as the various Higgs masses.
For $\alpha^{ij}\ll1$, we find
\begin{equation}
\theta^{ij}\approx\left({m_2\over 12m_1}\right)\alpha^{ij}\ ,
\label{mixing}
\end{equation}
with the exact relation depending on the complete Higgs spectrum.
{}From Eq.~(\ref{mixing}) it follows that $\theta^{12}\sim10^{-2}$ as
demanded by the MSW solution is compatible with $\alpha^{12}<10^{-5}$
if $m_2>10^4m_1$.  This inequality, together with the relationship
between $m_1y^2$ and $m_2v^2$, implies that $v < 10$GeV and suggests
that the top quark acquires most of its mass by a mechanism different
from the standard Yukawa couplings of the electroweak model. This was
hinted at in \cite{frampton} but the details of the mass generation
mechanism are still under investigation.

We now address the phenomenology of the cubic mass relation,
Eq.~(\ref{cubicmass}).  Since the constant $C$ given in
Eq.~(\ref{constant}) depends on the soft breaking terms and can be made
arbitrarily small, the cubic formula does not give any prediction for
the absolute masses.  Nevertheless, it gives definite predictions for
the ratios of neutrino masses.  Because of the cubic relation, the strongest
bound on $C$ comes from the $\nu_\tau$ mass bound, $m(\nu_\tau)<35$MeV.
Non-zero mixing angles $\theta^{ij}$
will perturb the exact cubic relation, but the deviations are small for
small mixing.

For $m(\nu_\tau)$ we may adopt the estimate $m(\nu_\tau)=29.3$eV from
the hot dark matter model of Sciama \cite{hdm}.  The massive $\nu_\tau$
is not stable, and will decay radiatively,
{\it e.g.}~by $\nu_\tau\to\nu_\mu\gamma$.  Since the gauge boson
contribution at one-loop is GIM suppressed \cite{loopdecay}, the
dominant decay mode is via the scalar exchange diagrams of
Fig.~\ref{fig1a} with a photon line attached in all possible ways
\cite{petcov}.  These one-loop diagrams give rise to a
($\nu_i\leftrightarrow\nu_j$) transition dipole moment
\begin{eqnarray}
\mu_{ij}&\approx&{e\alpha_2\over\pi\sin^2\beta}{\delta m^2\over M_W^2}
{m(\ell_i)^3\over M(H)^4}\theta^{ij}f(m(\ell_i),M(H))\nonumber\\
&\approx&2eC{m(\ell_i)^3\over M_W^2M(H)^2}\theta^{ij}f(m(\ell_i),M(H))
\nonumber\\
&\approx&2e{m(\nu_i)\over M(H)^2}\theta^{ij}f(m(\ell_i),M(H))\ ,
\end{eqnarray}
where we have used the cubic mass relation, Eq.~(\ref{cubicmass}), to
arrive at the last line.  This shows the intimate relation between the
radiatively generated Majorana mass and the transition moment.  In the
above, we have taken $L_i$ to be the heavier family and dropped terms
of order $m(\ell_j)/m(\ell_i)$.  $M(H)$ is a typical Higgs mass, and
the function $f$ is given by
\begin{equation}
f(m(\ell_i),M(H))\approx2+\ln\left({m(\ell_i)^2\over M(H)^2}\right)\ .
\end{equation}
Note that this gives a moderate enhancement to the transition moment since
typically $m(\ell)\ll M(H)$.

Assuming CP invariance, the transition dipole moment is either electric
or magnetic, depending on the CP eigenvalues of the neutrinos
\cite{transition}.  It gives rise to a decay width
\begin{equation}
\Gamma(\nu_i\to\nu_j\gamma)={1\over8\pi}m(\nu_i)^3|\mu_{ij}|^2\ ,
\end{equation}
assuming $m(\nu_i)\gg m(\nu_j)$.  For $\nu_\tau$, the
resulting radiative decay lifetime is given by
\begin{equation}
\tau\approx {1\over2\alpha}{M(H)^4\over m(\nu_\tau)^5}
{1\over(\theta^{23})^2} {1\over f^2(m(\tau),M(H))}\ ,
\end{equation}
giving lifetimes as short as $\sim 10^{23}$s as required by the
cosmological scenario of Ref.~\cite{hdm}.  Using the charged lepton
masses then gives from Eq.~(\ref{cubicmass}) $m(\nu_\mu)=6.2\times
10^{-3}$eV and $m(\nu_e)=6.9\times10^{-10}$eV.  These neutrino masses
are well below current experimental limits and are nicely consistent
with the MSW mechanism \cite{msw}.

This large scalar enhancement to the $\nu_\tau$ decay rate also occurs
in the charged leptonic sector for radiative processes such as $\mu\to
e\gamma$.  Although this is strongly constrained, the smallness of
$m(\nu_\mu)$ from the cubic formula gives a predicted branching ratio
well under current experimental limits.  Thus bounds on the tree level
dilepton contribution to $\mu\to3e$ remain the strongest limit on lepton
family mixing in the 331 model.

The effective Majorana mass of $\nu_e$, $\langle m(\nu_e)\rangle=\sum_i
m(\nu_i) (\delta^{1i}+\theta^{1i})^2$, leads to a contribution to
neutrinoless double
beta decay, $(\beta\beta)_{0\nu}$.  Assuming $\theta^{13}<10^{-2}$ since
only $m(\nu_\tau)$ is significant, this
contribution is at least three orders of magnitude
below the current experimental limit \cite{double}.  In the 331 model, there
are additional contributions to $(\beta\beta)_{0\nu}$ such as those
indicated in the diagrams of Fig.~\ref{fig3}.  We estimate, however, that
these dilepton diagrams are also many orders of magnitude beyond current
empirical limits.

We have seen that the minimal 331 model is easily extended to give a
highly predictive framework for generating neutrino masses and mixing.  The
basis for this lies in the fact that all leptons in a single family fall in
a single SU(3) triplet with mass generation coming from common Yukawa
couplings.  Our principal result is the cubic mass relation,
Eq.~(\ref{cubicmass}).  It provides a consistent relation between
$m(\nu_\tau)$ as required for hot dark matter and $m(\nu_\mu)$ as
needed for the solar neutrino deficit.  It further predicts $m(\nu_e)$
of about $0.7\times 10^{-9}$eV.

The derivation we have given is flawed only by our treatment of the
SU(2) triplet Higgs.  A fine tuning was necessary to prevent the
neutral component, $H^{22}$, from picking up a VEV and hence giving the
neutrinos mass at tree-level.  If $\langle H^{22}\rangle$ is
non-vanishing, the tree-level neutrino masses are of the form
$m(\nu_i)\approx m(\ell_i) \delta m^2/4M(H^{22})^2$, assuming the
triplet VEV takes on its natural scale.  This then dominates the cubic
term on the right hand side of Eq.~(\ref{cubicmass}) and leads to a
linear mass relation which allows for a simultaneous solution in terms of
neutrino oscillations of both the solar and atmospheric neutrino puzzles.
Nevertheless, finding a model like \cite{frampton,pleitez} but where the
neutral
component of the triplet scalar is either absent or has a vacuum value
constrained to vanish by a symmetry might be a fruitful line to pursue to
make a rigorous derivation of a cubic mass relation.

We finally mention that a cubic mass relation can be accommodated by the
see-saw mechanism using an inverse mass hierarchy for the heavy singlet
Majorana masses \cite{harari}.  However, the
present cubic mass relation, Eq.~(\ref{cubicmass}), is not generated by a
see-saw mechanism and does not involve right-handed neutrino fields.
This relation comes instead because the 331 model provides a common
framework for both charged leptons and neutrinos, once again showing the
highly predictive aspects of this theory.

This work was supported in part by the U.S.~Department of Energy under Grant
No.~DE-FG05-85ER-40219.

\begin{figure}
\caption{One-loop diagrams contributing to neutrino mass.  In addition,
there are four more diagrams with the charged lepton replaced by its charge
conjugate.}
\label{fig1a}
\end{figure}
\begin{figure}
\caption{Some of the potentially most significant additional contributions to
neutrinoless double beta decay in the 331 model.}
\label{fig3}
\end{figure}

\end{document}